\documentstyle[aps,amssymb,amsfonts,epsf,twocolumn]{revtex}

 \def\idty{\openone}
 \def\tr{{\rm tr}}
 \def\ketbra#1#2{\vert#1\rangle\!\langle#2\vert}
 \def\Cx{{\mathbb C}}\def\H{{\cal H}}\def\K{{\cal K}}

\title{\bf Counterexample to an additivity conjecture for output purity of quantum channels}
 
\author{R.~F. Werner\thanks{Electronic mail: r.werner@tu-bs.de}}
\address{
Institut f\"ur Mathematische Physik, TU Braunschweig,
Mendelssohnstr.3, 38106 Braunschweig, Germany.
 }

\author{and A.~S. Holevo\thanks{Electronic mail: holevo@mi.ras.ru}} \address{ Steklov Mathematical Institute,
Gubkina 8, 117966 Moscow, Russia}

\begin{document}
\maketitle
%%%%%%%%%%%%%%%%%%%%%%%%%%%%%%%%%%%%%%%%%%%%%%%%%%%%%%%%%%%%%%%%%%%%
\narrowtext
%%%%%%%%%%%%%%%%%%%%%%%%%%%%%%%%%%%%%%%%%%%%%%%%%%%%%%%%%%%%%%%%%%%%%%%%%%%
\begin{abstract}
A conjecture arising naturally in the investigation of additivity
of classical information capacity of quantum channels states that
the maximal purity of outputs from a quantum channel, as measured
by the $p$-norm, should be multiplicative with respect to the
tensor product of channels. We disprove this conjecture for
$p>4.79$. The same example (with $p=\infty$) also disproves a
conjecture for the multiplicativity of the injective norm of
Hilbert space tensor products.
\end{abstract}

\section{Statement of the problem}
\subsection{Multiplicativity of output purity}
In many applications of quantum information theory the
entanglement of states or the capacity of channels appear as
resources, which are needed to perform a task and are used up in
the process. Therefore, it is natural to expect that certain
entanglement or capacity measures should be {\it additive} in the
sense that preparing two pairs of entangled particles should give
us twice the entanglement of one pair and, similarly, using a
channel twice doubles its capacity. However, such additivity
properties have turned out to be notoriously difficult to prove,
and in some cases folk conjectures claiming additivity have turned
out to be wrong.

The purpose of this note is to provide a counterexample of this
kind i.e. to show that a family of quantities, which had been
conjectured to be additive in an earlier paper by the present
authors \cite{AHW}, actually is not. The quantities considered all
characterize the highest purity of the outputs of a channel. That
is, if $S$ is a completely positive map, taking density operators
on a finite dimensional Hilbert space $\H_1$ to density operators
on another finite dimensional Hilbert space $\H_2$, and
$1<p<\infty$,  we defined
\begin{equation}\label{nu}
 \nu_p(S)=\sup \Vert S(\rho)\Vert_p\ ,
\end{equation}
where the supremum is over all input density operators, and $\Vert
\rho\Vert_p=(\tr \vert\rho\vert^p)^{1/p}$ is the standard
$p$-norm. The conjecture in \cite{AHW} was that $\log \nu_p$ is
additive in the sense that
 \begin{equation}\label{addi}
  \nu_p(S_1\otimes S_2)=\nu_p(S_1)\nu_p(S_2)
\end{equation}
for arbitrary channels $S_1,S_2$.

This conjecture was supported by some numerical evidence (the
inequality ``$\geq$'' being trivial anyhow), and a proof for very
noisy and almost noiseless channels in \cite{AHW}, as well as some
depolarizing channels. Further supporting evidence was given by
C.~King \cite{King1,King2}. The main application would probably
have been in the limit $p\to1$, where it would be the additivity
of ``maximal purity as measured by entropy''. This in turn is
closely related \cite{AHW,King2,shor} to the question of additivity of
classical channel capacity, i.e., whether transmission of
classical information over multiple quantum channels can sometimes
be improved by using entangled signal states.

\subsection{Injective tensor norm for Hilbert space vectors}
The case $p=\infty$, i.e., when $\Vert\cdot\Vert_\infty$ is the
ordinary operator norm, is implied by another additivity
conjecture, namely for the {\it injective tensor norm} of Hilbert
space vectors. For any vector
$\Phi\in\H_1\otimes\cdots\otimes\H_N$  we define this norm as
\begin{equation}\label{nui}
  \mu_N(\Phi)=\sup\vert\langle \Phi,
                \phi_1\otimes\cdots\otimes\phi_N\rangle\vert\ ,
\end{equation}
where the supremum is over all tuples of vectors
$\phi_\alpha\in\H_\alpha$ with $\vert\phi_\alpha\vert=1$. The
conjectured property for this quantity was that
$\mu_N(\Phi\otimes\Psi)=\mu_N(\Phi)\mu_N(\Psi)$, where
$\Phi\in\H_1\otimes\cdots\otimes\H_N$ and
$\Psi\in\K_1\otimes\cdots\otimes\K_N$ may be in different $N$-fold
Hilbert space tensor products, and the supremum in
$\mu_N(\Phi\otimes\Psi)$ is taken over unit vectors
$\phi_\alpha\in\H_\alpha\otimes\K_\alpha$. Again, the inequality
``$\geq$'' and the case $N=2$ are trivial. The connection with the
previous problem is seen by writing $S$ in Kraus form
$S(\rho)=\sum_xA_x\rho A_x^*$. Then
\begin{eqnarray}\label{mu2nu}
  \bigl\langle\varphi,S(\ketbra\phi\phi)\varphi\bigr\rangle
   &=& \sum_x \bigl\vert\langle\Phi,A_x\phi\rangle\bigr\vert^2
             \nonumber\\
   &=&\sup_\psi \Bigl\vert
        \sum_x\overline{\psi_x}\langle\varphi,A_x\phi\rangle\Bigr\vert^2\;,
\end{eqnarray}
where at the last equality we consider the
$\langle\varphi,A_x\phi\rangle$ as the components of a Hilbert space
vector, whose norm can also be written as the largest scalar
product with a unit vector. Accordingly, the supremum in the last
line is over all unit vectors $\psi$. Taking the supremum over the
unit vectors $\phi$ and $\varphi$, too, we find that
\begin{equation}\label{mu2nu2}
  \nu_\infty(S)=\mu_3(\widetilde A)^2\;,
\end{equation}
 where $\widetilde A$ denotes the vector in a
threefold Hilbert space tensor product with components $\langle
h_j ,A_xe_k\rangle$, where $h_j$ and $e_k$ are
orthonormal bases of the appropriate spaces. In particular, since
the tensor product of channels $S$ corresponds to the tensor
product of vectors $\widetilde A$, the conjectured
multiplicativity of $\mu_3$ would imply the multiplicativity of
$\nu_\infty$. Conversely, the counterexample given to the latter
disproves the multiplicativity of $\mu_N$ for all $N\geq3$.

\section{The counterexample}
We give an explicit example of a channel violating conjecture
(\ref{addi}) for large values of $p$. It is the channel $S$ on the
$d\times d$-matrices defined as
\begin{eqnarray}\label{S}
  S(\rho)&=& \frac1{d-1}\Bigl(\tr(\rho)\idty-\rho^T\Bigr)\\
    &=&\frac1{2(d-1)}\sum_{ij}\bigl(\ketbra ij-\ketbra ji\bigr)^*\rho
            \bigl(\ketbra ij-\ketbra ji\bigr)\;.
\end{eqnarray}
Here $\rho^T$ denotes the matrix transpose with respect to some
fixed basis. In the first form it is easy to verify that $S$ is
linear and trace preserving, in the second it becomes clear that
it is also completely positive. The equality of the two forms is a
straightforward exercise. Further interesting properties are that
$S$ is hermitian with respect to the Hilbert-Schmidt scalar
product $(A,B)\mapsto\tr(A^*B)$ on $d\times d$-matrices, and
covariant for arbitrary unitary transformations in the sense that
 $S\circ {\rm ad}_U={\rm ad}_{\overline{U}}\circ S$, ${\rm
ad}_U(X)=UXU^*$, where $\overline{U}$ denotes the matrix
element-wise complex conjugate of a unitary in the fixed basis. We
remark that $S$ is the dual of the state which provided the
counterexample to the additivity of the relative entropy of
entanglement for bipartite states in \cite{NOaddER}, i.e., that
state, the normalized projection onto the Fermi subspace of
$\Cx^d\otimes \Cx^d$ ($d\geq3$), is obtained by acting with $S$ on
one partner of a maximally entangled pair on $\Cx^d\otimes \Cx^d$.

Now  $\rho\mapsto\Vert S(\rho)\Vert_p$ is a convex function, and
hence takes its maximum on the extremal states. Therefore it
suffices to take pure input states, for which
\begin{equation}\label{pur1}
  S(\ketbra \phi\phi)
    =\frac1{d-1}\Bigl(\idty-\ketbra{\overline{\phi}}{\overline{\phi}}
                \Bigr)\;.
\end{equation}
Clearly, the $p$-norm is the same for all pure inputs, and we get
\begin{equation}\label{nupS}
  \nu_p(S)=(d-1)^{-(1-1/p)}\;.
\end{equation}

On the other hand, let us consider $S\otimes S$ acting on a pure
state $\Phi$. Due to covariance we may take $\Phi$ in Schmidt
diagonal form $\Phi=\sum_\alpha c_\alpha\vert\alpha\alpha\rangle$.
Then with the reduced density operator $\rho=\sum_\alpha
c_\alpha^2 \ketbra\alpha\alpha$ we get
\begin{eqnarray}\label{SSPhi}
  S\otimes S&&(\ketbra\Phi\Phi)=   \nonumber\\
              &&=\frac1{(d-1)^2}\Bigl(
       \idty-\idty\otimes\rho-\rho\otimes\idty +
          \ketbra\Phi\Phi\Bigr)\;.
\end{eqnarray}
 We now specialize further to {\it maximally entangled} $\Phi=\Phi_m$, i.e.,
all $c_\alpha=1/\sqrt d$. and $\rho=(1/d)\idty$, all terms in this
expression commute, and the operator in parenthesis has one
eigenvalue $(1-2/d)$ with multiplicity $(d^2-1)$ and a
non-degenerate eigenvalue $(1-2/d+1)$. From this we find, for
$d=3$:
\begin{equation}\label{SSPhi-pnorm}
  \Vert S\otimes S(\ketbra{\Phi_m}{\Phi_m})\Vert_p
    =\frac13\bigl(1+2^{3-2p}\bigr)^{1/p}\;.
\end{equation}
If additivity were true, the following quantity should be
negative, becoming zero upon maximization with respect to $\Phi$ :
\begin{equation}
 \Delta(p,\Phi)=\log\Vert S\otimes S(\ketbra\Phi\Phi)\Vert_p-2\log\nu_p(S)
\end{equation}
However, inserting for $\Phi$ the maximally entangled state we get
\begin{equation}\label{del}
   \Delta(p,\Phi_m)=\log\frac43+\frac1p\log\left(\frac14+2^{1-2p}\right).
\end{equation}
This quantity is plotted in Fig.~1. Thus additivity is violated
for $p>p_0$, where the zero $p_0$ is numerically determined as
$p_0=4.7823$. In particular, we get
$\Delta(\infty,\Phi_m)=\log(4/3)>0$, so additivity fails for
$p=\infty$. Transcribed to the problem for injective Hilbert space
norms, the counterexample is the unique antisymmetric vector
$\Phi$ on $\Cx^3\otimes\Cx^3\otimes\Cx^3$, tensored with itself.
On the other hand, for $p\to1$ we have $\Delta(p,\Phi_m)<0$, so in
this case, which is of main interest for classical channel
capacity, the additivity conjecture survives.

\begin{figure}[hbt]
\begin{center}
    \leavevmode
    \epsfysize=4.5cm
    \epsfbox{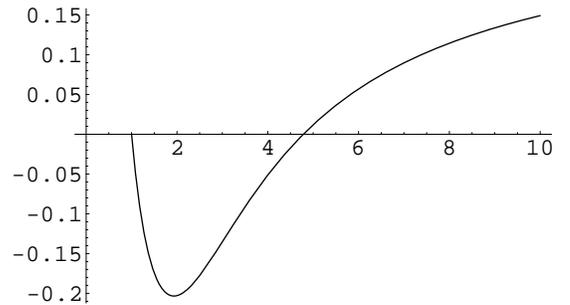}
    \caption{ $\Delta(p,\Phi_m)$ over $p$, as explained in the text.  }
\end{center}
\end{figure}

The boundary point $p_0$ cannot be improved by choosing another
vector $\Phi$, i.e., the maximizing $\Phi$ in the definition of
$\nu_p(S\otimes S)$ jumps discontinuously from a product state to
a maximally entangled state, as $p$ increases beyond $p_0$. This
is seen by plotting $\Delta$ for fixed $p$ over the Schmidt
parameters of $\Phi$. There is little difference between the plots
for $p=4$ and $p=5$ except that the global maximum switches from
corners ($\Phi$ a product vector) to the center ($\Phi=\Phi_m$).

\begin{figure}[hbt]
\begin{center}
    \leavevmode
    \epsfysize=4cm \epsfbox{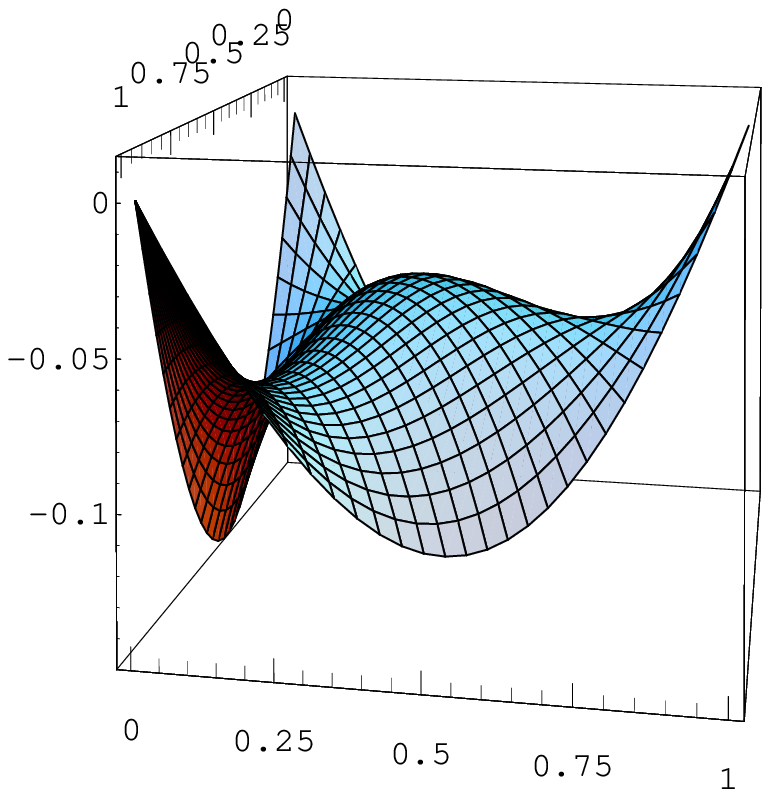}
    \epsfysize=4cm \epsfbox{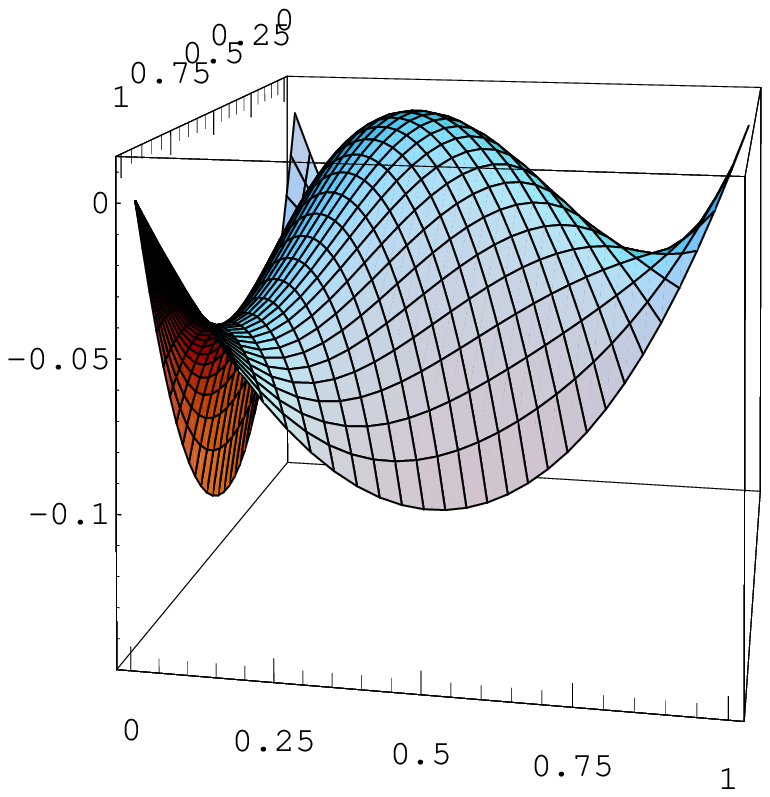}
    \caption{ $\Delta(p,\Phi)$ for $d=3$ over the
         Schmidt parameters $c_1^2$ and $c_2^2$ of $\Phi$,
         for $p=4$ (left) and for $p=5$ (right).  }
\end{center}
\end{figure}

\section{Concluding Remarks}
We have seen that multiplicativity of maximal purity depends
crucially on how we measure purity: if we use $p$-norms for large
$p$, it fails. But the conjecture remains open for small $p$, and
in particular for purity as measured by entropy. There seems to be
no simple modification of the example given here to improve the
critical value of $p$. On the other hand, it is hardly expected to
be optimal.

\acknowledgments{R.F.W. acknowledges correspondence with K. Floret
on injective tensor norms and financial support from the DFG and
the EU (IST-1999-11053). A.S.H. acknowledges support from INTAS
project 00-738. }

\end{document}